\documentclass[12pt,thmsa]{article}
\usepackage{amssymb,epsfig}

\input{tcilatex}
\begin{document}

\newpage \setcounter{page}{0}
\begin{titlepage}
\begin{flushright}
Berlin Sfb288 Preprint \\
quant-ph/9707059
\end{flushright}
\vspace{0.5cm}
\begin{center}
{\Large {\bf Ionization Probabilities through ultra-intense 
Fields in the extreme Limit} }

\vspace{1.8cm}
{\large A. Fring$^*$, V. Kostrykin$^{\dag}$ and R. Schrader$^*$}
\footnote{e-mail addresses: Fring@physik.fu-berlin.de,

Kostrykin@ilt.fhg.de

Schrader@physik.fu-berlin.de} \\
\vspace{0.5cm}
{\em $^*$ Institut f\"ur Theoretische Physik\\
Freie Universit\"at Berlin, Arnimallee 14, D-14195 Berlin, Germany\\
$^{\dag}$  Lehrstuhl f\"ur Lasertechnik\\
RWTH Aachen, Steinbachstr. 15, D-52074 Aachen, Germany}
\end{center}
\vspace{1.2cm}

\renewcommand{\thefootnote}{\arabic{footnote}}
\setcounter{footnote}{0}

\begin{abstract}
We continue our investigation concerning the question of whether
atomic bound states begin to stabilize in the ultra-intense field
limit. The pulses considered  are essentially arbitrary, but we
distinguish between three situations. First the total classical
momentum transfer is non-vanishing, second not both the total
classical momentum transfer and  the total classical displacement
are vanishing together with the requirement that the potential has
a finite number of bound states and third both the total classical
momentum transfer and  the total classical displacement are
vanishing. For the first two cases we rigorously prove, that the
ionization probability tends to one when the amplitude of the pulse
tends to infinity and the pulse shape remains fixed. In the third
case the limit is strictly smaller than one. This case is also related
to the high frequency limit considered by Gavrila et al.
\end{abstract}
\centerline{July 1997}
\end{titlepage}
\newpage

Ionization probabilities of atomic systems in the presence of
intense laser fields are in general poorly predicted. Intense means
here that the field intensities are of comparable size in magnitude
with the ionization energy of the potential and hence conventional
perturbation theory ceases to be valid. Numerous different methods
for theoretical investigations have been carried out in order to
treat the new intensity regime, such as perturbative methods around
the Gordon-Volkov solution \cite{GordonVolkov} of the
Schr\"{o}dinger equation
\cite{Keld,GeltT,Reiss,Gelt1,Grobe,Gelt2,Gelt3}, fully numerical
solutions of the Schr\"{o}dinger equation \cite
{Col,Bard,LaG,SuEb,Burnett,Dorr,PO,FDS}, Floquet solutions \cite
{Floquet,Pot,Faisal}, high frequency approximations \cite{HFA} or
analogies to classical dynamical systems \cite{classical}. Some of
these investigations have led to the prediction of so-called atomic
stabilization, which means that the ionization probability is
supposed to decrease once a certain critical intensity has been
surpassed. However, several authors have raised doubts and question
whether such an effect really exists \cite
{Gelt2,Chen,Krain,Gelt1,Gelt3,FKS,FFS,KS}. For reviews on the
subject we refer the reader to \cite{BRK,KulS,KrainDel,Gelt4}.

In this paper we want to continue our previous investigations
\cite{FKS,FFS,KS} and answer the question concerning 
the ionization probability in the limit when
the field amplitude tends to infinity, while the pulse shape
remains fixed. Of course strictly speaking one would have to include
relativistic effects into the analysis at some high intensities and
then a proper quantum field theoretical treatment is needed.
However, the Schr\"{o}dinger theory with the a.c. Stark Hamiltonian
is consistent in itself, also
in that regime, and in this light the limit becomes meaningful.
Clearly our analysis does not capture the effect of window
stabilization, which is the purported phenomenon that stabilization
only occurs in a certain regime of high intensities and then the
ionization probability tends to one once this regime is surpassed.

We consider the Schr\"{o}dinger equation involving some potential $V\left(
\vec{x}\right) $, for instance the atomic potential, coupled to a classical
linearly polarized electric field in the dipole approximation $E\left(
t\right) $
\begin{equation}
i\frac{\partial \psi \left( \vec{x},t\right) }{\partial t}=\left( -\frac{
\Delta }{2}+V\left( \vec{x}\right) +z\cdot E\left( t\right) \right) \psi
\left( \vec{x},t\right) =H\left( t\right) \psi \left( \vec{x},t\right).  \label{Schro}
\end{equation}
We use atomic units $\hbar =e=m_{e}=c\alpha =1$ and we will mainly
adopt the notations in \cite{FKS}. We now want to state precisely
which type of potentials and electric fields are included in our
analysis.

\begin{description}
\item[Assumptions on V:]  $V\left( \vec{x}\right) $\textit{\ is a real
measurable function on } $\mathbb{R}^{3}$. \textit{\ To each } $\varepsilon >0$
\textit{\ one may decompose }$V$\textit{\ as}
\begin{equation}
V=V_{1}+V_{2}  \label{AssV1}
\end{equation}
\textit{where }$V_{1}$\textit{\ is in }$L^{2}\left( \mathbb{R}^{3}\right) $
\textit{\ (i.e. square integrable) with compact support and }$V_{2}$\textit{
\ is in }$L^{\infty }\left( \mathbb{R}^{3}\right) $\textit{\ with}
\begin{equation}
\left\| V_{2}\right\| _{\infty }=\text{ess}\sup\limits_{\vec{x}\in
\mathbb{R}^{3}}\left| V_{2}(\vec{x})\right| \leq\varepsilon \quad .
\label{AssV2}
\end{equation}
\textit{Furthermore we assume that }$H=H_{0}+V$ with $H_{0}=-\frac{\Delta }{2
}$\textit{\ has no positive bound states.}
\end{description}

Relation (\ref{AssV2}) means that up to a set of measure zero
$V_{2}(\vec{x})$ is bounded in absolute value by $\varepsilon $. We
note that the potentials of atoms or molecules arising from Coulomb
pair interactions belong to this wide class. To obtain for instance the
decomposition (\ref{AssV1}) for the Coulomb potential $1/|\vec{x}|$
we set
\begin{displaymath}
\frac{1}{|\vec{x}|}=\frac{\chi_{1/\varepsilon}(\vec{x})}{|\vec{x}|}+
\frac{1-\chi_{1/\varepsilon}(\vec{x})}{|\vec{x}|},
\end{displaymath}
where $\chi_R(\vec{x})$ is the characteristic function 
of the ball $\left\{\vec{x}:\ |\vec{x}|\leq R
\right\}$ of radius $R$,
\begin{displaymath}
\chi_R(\vec{x})=\left\{\begin{array}{ccc}
                       1 & \mbox{for} & |\vec{x}|\leq R \\
                       0 & \mbox{for} & |\vec{x}|> R \;\; .
                       \end{array}\right.
\end{displaymath}

Potentials satisfying the above assumptions are Kato small, i.e.
for each  $\alpha $ with $0<\alpha \leq 1,$ there exists a
constant $\beta = \beta(\alpha)
\geq 0$ such that
\begin{equation}
\left\| V\psi \right\| \leq \alpha \left\| -\frac{\Delta }{2}\psi \right\|
+\beta \left\| \psi \right\|  \label{Katos}
\end{equation}
for all $\psi \in $\textit{\ }$L^{2}\left( \mathbb{R}^{3}\right) $
with $\Delta \psi \in $\textit{\ }$L^{2}\left( \mathbb{R}^{3}\right) $. 
The Hamiltonian $H$ is
self-adjoint on the Hilbertspace $L^{2}\left( \mathbb{R}^{3}\right) $ and the domains
$\mathcal{D}\left( H\right) $ and $%
\mathcal{D}\left(H_{0}\right) $ of definition of $H$ and $H_{0}$ agree \cite
{Kato}. $H$ is bounded from below and has no positive eigenvalues if $V$ decays suitably
at infinity \cite{RS,FR}.

As for the conditions on the electric field, we assume that it takes on the
form
\begin{equation}
E\left( t\right) =E_{0}f\left( t\right) ,
\end{equation}
where $f\left( t\right) $ is assumed to be measurable in $t$ with $f\left(
t\right) =0$ unless $0\leq t\leq \tau .$ We call $\tau $ the pulse duration,
$f(t)$ the pulse shape
and $E_{0}$ the amplitude of the pulse $E\left( t\right) .$ Further we
introduce the quantities
\begin{eqnarray}
b\left( t\right) &=&\int\limits_{0}^{t}ds\,E\left( s\right)
=E_{0}\int\limits_{0}^{t}ds\,f\left( s\right) =E_{0}b_{0}\left( t\right) \\
c\left( t\right) &=&\int\limits_{0}^{t}ds\,b\left( s\right) =tb\left(
t\right) -\int\limits_{0}^{t}ds\,E\left( s\right) s=E_{0}c_{0}\left( t\right)
\;\; .
\end{eqnarray}
With $e_{z}$ being the unit vector in $z$-direction, $b\left( \tau \right)\ e_{z}$ is the
total classical momentum transfer and $c\left( \tau \right) e_{z}$ the total classical
displacement. We are now in the position to formulate more precisely our assumptions on
the electric field, that is on the pulse shape $f(t)$.

\begin{description}
\item[Assumptions on E: ]  $f\left( t\right) $\textit{\ is a real measurable
non-vanishing function in }$t,$\textit{\ with support in the interval }$
\left[ 0,\tau \right] $\textit{\ such that }
\begin{equation}
b_{0}\left( \tau \right) \neq 0.  \label{AssE1}
\end{equation}
\textit{In case the potential possesses a finite number of bound states we
only assume that }
\begin{equation}
b_{0}\left( \tau \right) ^{2}+c_{0}\left( \tau \right) ^{2}\neq 0.
\label{AssE2}
\end{equation}
\textit{Finally, }$c_{0}\left( t\right) $\textit{\ is supposed to be piecewise
continuous possibly with a finite number of zeros in} $[0,\tau]$.
\end{description}

Of course the restrictions of a finite number of bound states excludes the Coulomb
Potential. However, we would like to remark that in general most numerical calculations in
this context implicitly also assume a finite number of bound states. When projecting on
bound states numerically, one is always forced to introduce a cut-off. Hence our analysis
allows also in that case a direct comparison with such computations. The gain in the
latter case is that when the requirement (\ref{AssE1}) is relaxed to (\ref{AssE2}) it
allows to include more types of pulses such as Gaussian etc.
All  pulses used in the
literature satisfy  assumption E. 

The ionization probability for any given normalized bound state
$\psi $ of the Hamiltonian $H$ is given by
\begin{equation}
\mathcal{P}\left( \psi \right) =\left\| \left( 1-P\right) U\left( \tau
,0\right) \psi \right\| ^{2}=1-\left\| PU\left( \tau ,0\right) \psi \right\|
^{2}\quad .
\end{equation}
Here $U\left( t^{\prime  },t \right) $ denotes the unitary 
time evolution operator from time $t$ to time
$t'$ associated to $H(t)$. Its existence
\footnote{In order to show the existence 
one  actually has to make some additional sufficient  assumptions on V,
namely one assumes  that $V_i(\vec{x} -u e_z)$, $i=1,2$ are 
$L^2$ and
$L^\infty$ valued continuous functions in $u$, such that
in addition
\begin{displaymath}
W_i(u)=\frac{\partial}{\partial u}V_i(\vec{x}-u e_z),\ i=1,2
\end{displaymath}
exists and satisfies $\|W_1(u)\|_p<\infty$ for some
$6/5<p\leq 4/3$ and $\|W_2(u)\|_\infty <\infty$
uniformly in $u$ on compact sets in $\mathbb{R}$. So strictly speaking
we have to extend our assumptions on V. However, for standard potentials
like Coulomb etc. this additional assumption is always satisfied and
we therefore omitted it above for the sake of clarity.}
follows from results 
in \cite{Yajima,Neidhardt} (for details see \cite{KoSch}).

Further $P$ denotes the orthogonal projection in
$L^{2}\left( \mathbb{R}^{3}\right) $ on the space spanned by the bound states of
$H=H_{0}+V$. For more details on the precise definition of the ionization probability and its properties we
refer the reader to \cite{FKS}. In what follows $f(t)$ and $V$ will be fixed.

We now formulate the Main Theorem of this article:

\begin{theorem}
With the above assumptions on the electric field E and the potential V, the ionization
probability $\mathcal{P}$ for any bound state $\psi $ of $ H=H_{0}+V$ tends to one for the
field amplitude $E_{0}$ going to infinity
\[
\lim_{\left| E_{0}\right| \rightarrow \infty }\mathcal{P}\left( \psi \right)
=1\quad .
\]
\end{theorem}

This improves a previous result in \cite{FKS} (see relation (3.31) therein), which stated
that $\lim_{\left| E_{0}\right| \rightarrow \infty }\mathcal{P}
\left( \psi \right) \geq 1-\tau ^{2}c$, where $c$ is a constant depending on
the potential $V$ and on $\psi$ only. The proof of this main theorem in
case of condition (\ref{AssE2}) shows that the finite dimensional 
projector $P$ may in fact be chosen arbitrarily. In particular in the case
of the Coulomb potential, $P$ may be the projector on the space spanned
by {\it any} finite set of bound states.

\noindent \textbf{Proof of the Main Theorem:}

\noindent To start the proof of the main theorem
following \cite{FKS} we may first rewrite the ionization
probability as
\begin{equation}
\mathcal{P}\left( \psi \right) =1-\left\| P\exp -ib\left( \tau \right)
z\cdot \exp ic\left( \tau \right) p_{z}\cdot U^{\prime }\left( \tau
,0\right) \psi \right\| ^{2}.
\end{equation}
Here $U^{\prime }\left( t^{\prime  },t\right) $ is the
unitary time evolution operator associated to  
the Stark Hamiltonian (\ref{Schro}) in
the Kramers-Henneberger gauge \cite{K,H}
\begin{equation}
H^{\prime }(t)=-\frac{\Delta }{2}+V\left( \vec{x}-c\left( t\right)
e_{z}\right) \quad .
\end{equation}

\vspace{5mm}
Crucial for the proof of the main theorem will be the next result, that in 
the limit $E_0 \rightarrow \infty $ the time evolution $U'$ for $H'(t)$
is just the free time evolution. We will need this result in the following 
form
\vspace{5mm}

\begin{theorem} For all $\varphi\in L^{2}\left( \mathbb{R}^{3}\right)  $
\begin{equation}\label{S13}
\lim_{|E_0|\rightarrow\infty}\left\|\left(U'(\tau,0)-
\exp -i\tau H_0\right)\varphi\right\|=0,
\end{equation}
i.e. $U'(\tau,0)$ converges strongly to $\exp (-i\tau H_0)$ as 
$|E_0|\rightarrow\infty$.
\end{theorem}

\vspace{5mm}

The proof of this theorem will proceed in several steps. Before we
begin with the proof we note that this is essentially  Kato's
theorem on the strong convergence of propagators for time dependent
Hamiltonians \cite{Kato2}. However we cannot use this theorem
directly since it is not valid for  Hamiltonians with Coulomb
interaction. Since $\left\|\left(U'(\tau,0)-\exp -i\tau H_0\right)
\varphi\right\| \leq 2 \left\| \varphi  \right\| $  it suffices to prove
(\ref{S13}) for all $\varphi\in\mathcal{D}(H_0)=\mathcal{D}(H)$, which is a 
dense set in $L^{2}\left( \mathbb{R}^{3}\right) $.

First we use Du Hamel's formula to write
\begin{eqnarray}\label{S14}
\left( U'(\tau,0) -\exp -i\tau H_0 \right) \varphi 
= -i \int_0^\tau
U'(\tau,s) V(\vec{x}-c(s)e_z)\exp -is H_0\ \varphi ds
\end{eqnarray}
with $\varphi\in\mathcal{D}(H_0)$. We note that by the spectral theorem 
$\exp -isH_0$ leaves $\mathcal{D}(H_0) $ 
invariant. Therefore from (\ref{S14}) it follows
that
\begin{equation}\label{S15}
\left\|\left(U'(\tau,0)-\exp -i\tau H_0\right)\varphi\right\| \leq
\int_0^\tau\left\| V(\vec{x}-c(s)e_z)\exp -isH_0 \cdot
\varphi\right\| ds,
\end{equation}
uniformly in $E_0$.

To proceed further we use the following

\vspace{5mm}

\begin{lemma}
\textit{For any }$\varphi \in \mathcal{D}\left( H\right) =\mathcal{D}\left(H_{0}\right)$
and all $s\in[0,\tau]$ with $c_0(s) \neq 0 $   \textit{\ one has }
\begin{equation}
\lim_{E_0 \rightarrow \infty }\left\| V\left( \vec{x}
-c(s) \gamma e_{z}\right) \varphi \right\| =0.  \label{Lem1}
\end{equation}
\end{lemma}

\vspace{5mm}

\begin{description}
\item[Proof:] It suffices to show that
for any $\varphi \in \mathcal{D}\left( H\right) =\mathcal{D}\left( H_{0}\right)$
\begin{displaymath}
\lim_{\left| \gamma \right| \rightarrow \infty }\left\| V\left( \vec{x}
-\gamma e_{z}\right) \varphi \right\| =0.
\end{displaymath}
We show that for arbitrary small $\varepsilon>0$  the estimate
$\|V(\vec{x}-\gamma e_z)\varphi\|<\varepsilon$ holds for all sufficiently 
large $\gamma>0$.

Since $V$ is Kato small and since $-\Delta $ commutes with translations, 
the potential in
the Kramers-Henneberger gauge satisfies a similar estimate
\begin{equation}\label{Kato-KH}
\left\| V\left( \vec{x}-\gamma e_{z}\right) \varphi \right\| \leq \left\| -
\frac{\Delta }{2}\varphi \right\| +\beta \left\| \varphi \right\| \quad 
\end{equation}
with fixed $\beta < \infty$ and for all $\gamma$.

Indeed,
\begin{eqnarray*}
\| V(\vec{x}-\gamma e_z)\varphi\| = \|V(\vec{x}) \exp (i\gamma p_z) \; 
\varphi \|\\
\leq \|H_0 \exp (i\gamma p_z) \; \varphi\| + 
\beta\| \exp (i\gamma p_z) \;\varphi\|=
\|H_0\varphi\| + \beta \|\varphi\|,
\end{eqnarray*}
where the last equality follows from the fact that $H_0$ commutes with the translations.
In comparison with (\ref{Katos}) we have taken $\alpha =1$ and chosen 
$ \beta = \max (\beta(\alpha = 1),1 )$. 
Hence it suffices to prove (\ref{Lem1})
on a core for $H_{0}$ which is also a core for $H$. We recall that $\mathcal{C}$ is a core
for a self-adjoint operator $A$ with domain $\mathcal{D}\left( A\right) $, if
$\mathcal{C}$ is contained and dense in $\mathcal{D}\left( A\right) $ with respect to the
topology in $\mathcal{D}\left( A\right) $ given by the norm $\left\| \varphi \right\|
_{\mathcal{D}\left( A\right) }=\left\| A\varphi \right\| +\left\| \varphi \right\| $.
Indeed, for a given $
\varphi \in \mathcal{D}\left( H_{0}\right) $ let $\varphi ^{\prime }\in $
$\mathcal{C}$ be such that
\[
\left\| H_{0}(\varphi -\varphi ^{\prime })\right\| +\left\| (\varphi
-\varphi ^{\prime })\right\| \leq \frac{\varepsilon }{2\beta }\;.
\]
Also let $\gamma (\varepsilon ,\varphi ^{\prime })$ be such that
\[
\left\| V\left( \vec{x}-\gamma e_{z}\right) \varphi ^{\prime }\right\| \leq
\frac{\varepsilon }{2}\qquad \hbox{for all}\quad \quad \gamma \geq \gamma
(\varepsilon ,\varphi ^{\prime }).
\]
Then
\begin{eqnarray*}
\left\| V\left( \vec{x}-\gamma e_{z}\right) \varphi \right\|  &\leq &\left\|
V\left( \vec{x}-\gamma e_{z}\right) (\varphi -\varphi ^{\prime })\right\|
+\left\| V\left( \vec{x}-\gamma e_{z}\right) \varphi ^{\prime }\right\|  \\
&\leq &\left\| H_{0}(\varphi -\varphi ^{\prime })\right\| +\beta \left\|
(\varphi -\varphi ^{\prime })\right\| +\frac{\varepsilon }{2}\leq
\varepsilon.
\end{eqnarray*}
Now $C_{0}^{\infty }\left(\mathbb{R}^{3}\right)$, the set of smooth functions on
$\mathbb{R}^{3}$ with compact support, is such a core and we will now prove (\ref{Lem1})
on this core. Assuming that $\varphi \in C_{0}^{\infty }\left(\mathbb{R}^{3}\right) $ is
normalized, we obtain with the assumptions on $V$
\begin{eqnarray*}
\left\| V\left( \vec{x}-\gamma e_{z}\right) \varphi \right\|  &\leq &\left\|
V_{1}\left( \vec{x}-\gamma e_{z}\right) \varphi \right\| +\left\| V_{2}\left(
\vec{x}-\gamma e_{z}\right) \varphi \right\|  \\ &\leq &\left\| V_{1}\left( \vec{x}-\gamma
e_{z}\right) \varphi \right\| +\varepsilon.
\end{eqnarray*}
For $\left| \gamma \right| $ sufficiently large $V_{1}\left( \vec{x}-\gamma e_{z}\right)
\varphi =0$ and the lemma follows.
\end{description}

\vspace{5mm}

We proceed with the proof of the theorem. Since for fixed 
$\varphi \in \mathcal{D}\left( H_{0}\right) $ the map $\exp(-isH_0)\ : 
\left[0,\tau \right]\rightarrow \mathcal{D}\left( H_{0}\right) $ given by
$s \mapsto \exp(-isH_0) \varphi  $
is continuous, the set $\mathcal{S}=\left\{ \exp(-isH_0) \varphi 
\right\}_{0 \leq s \leq \tau}$ 
is compact in $\mathcal{D}\left( H_{0}\right)$. Therefore
\begin{displaymath}
\|V(\vec{x}-c(s)e_z)\psi\| \rightarrow 0
\end{displaymath}
as $\left| E_0  \right| \rightarrow \infty$
uniformly in $\psi\in\mathcal{S}$ for  all $s\in[0,\tau]$ 
except the finite set where $c_0(s)=0 $  (see e.g.
\cite{Kato:book}). Now the r.h.s. of (\ref{S15}) for any $\varphi\in\mathcal{D}(H_0)$
can be bounded by
\begin{equation}\label{fastfertig}
\left\|\left(U'(\tau,0)-\exp -i\tau H_0\right)\varphi\right\| \leq
\int_0^\tau \sup_{\psi\in\mathcal{S}} \|V(\vec{x}-c(s)e_z)\psi\| ds.
\end{equation}

\noindent From (\ref{Kato-KH}) and the definition of $\mathcal{S}$
it follows also that
\begin{displaymath}
\|V(\vec{x} -c(s)e_z)\psi\| \leq C_\varphi
\end{displaymath}
for all $\psi \in \mathcal{S}$ and all $s \in \left[0,\tau \right]$
with
\begin{displaymath}
C_\varphi=\left\|-\frac{\Delta }{2}\varphi \right\| +\beta \left\| \varphi \right\|
\end{displaymath}
for every $\varphi\in\mathcal{D}(H_0)$ uniformly in $E_0$.

By the  Lebesgue dominated convergence theorem we therefore have that
the right hand side of (\ref{fastfertig}) tends to zero as 
$|E_0| \rightarrow \infty$.

This completes the proof of the theorem.

\vspace{5mm}

\begin{description}
\item[Remark 5]  From the preceding discussion, it is obvious how to weaken
the last condition in the assumptions on E. Assume we may divide 
the interval $\left[
0,\tau \right] $ into $2N+1$ parts as $0=\tau _{0}<\tau _{1}<\cdots <\tau
_{2N}<\tau _{2N+1}=\tau ,$ such that $c_{0}(t)$ vanishes identically in the
intervals $\left[ \tau _{2j},\tau _{2j+1}\right] $ $\left( 0\leq j\leq
N\right) $ and is non-zero except for a finite set  in the intervals
$\left[ \tau _{2j+1},\tau _{2j+2}\right] $ $\left( 0\leq j\leq N-1\right) $.
Then $U^{\prime }\left( \tau ,0\right) $ converges strongly to
\begin{equation}
U^{\prime \prime }=e^{-i\left( \tau _{2N+1}-\tau _{2N}\right) H_{0}}\cdot
e^{-i\left( \tau _{2N}-\tau _{2N-1}\right) H}\cdots e^{-i\left( \tau
_{2}-\tau _{1}\right) H}\cdot e^{-i\left( \tau _{1}-\tau _{0}\right)
H_{0}}\quad .  \label{REM}
\end{equation}
Since $f(t)$ is by assumption not identically zero, $c_{0}\left( t\right) $ is not
identically zero on $\left[ 0,\tau \right] $, so 
$ U^{\prime \prime } \neq \exp - i \tau H  $.
\end{description}

\noindent To prove the main theorem it suffices now by Theorem 2 and the
obvious estimate
\begin{eqnarray}
& &\left\| P e^{-ib\left( \tau \right) z }
\cdot e^{ic\left( \tau \right) p_{z}}\cdot U'(\tau,0) \psi
\right\| \nonumber  \\
&\leq & \left\| P e^{-ib\left( \tau
\right) z }\cdot e^{ ic\left( \tau \right) p_{z}}
\cdot e^{-i\tau H_{0}} \psi
\right\|  + \left\| P e^{ -ib\left( \tau
\right) z }\cdot e^{ ic\left( \tau \right) p_{z}}\cdot \left( 
 U'(\tau,0) -   e^{-i\tau H_{0}} \right) \psi
\right\| \nonumber \\
&\leq & \left\| P e^{-ib\left( \tau
\right) z }\cdot e^{ ic\left( \tau \right) p_{z}}
\cdot e^{-i\tau H_{0}} \psi
\right\|  + \left\| \left( U'(\tau,0) -   e^{-i\tau H_{0}} \right) \psi
\right\| \nonumber
\end{eqnarray}
to show that
\begin{equation}
\lim_{\left| E_{0}\right| \rightarrow \infty }\left\| P\exp -ib\left( \tau
\right) z\cdot \exp ic\left( \tau \right) p_{z}\cdot \exp -i\tau H_{0}\psi
\right\|  = 0  \,\,\,.  \label{con1}
\end{equation}
Here $\exp -i\tau H_{0}$ has to be replaced by $U^{\prime \prime }$ in case
remark 5 applies. Since $\exp -i\tau H_0$ leaves $\mathcal{D}\left(
H_{0}\right) $ invariant, it is enough to show
\begin{equation}
\lim_{\left| E_{0}\right| \rightarrow \infty }\left\| P\exp -ib\left( \tau
\right) z\cdot \exp ic\left( \tau \right) p_{z}\varphi \right\| = 0 \,\,\,.
\label{con2}
\end{equation}
for all $\varphi \in \mathcal{D}\left( H\right) =\mathcal{D}\left(
H_{0}\right) $ in order to prove (\ref{con1}).

We now modify some arguments already used in \cite{EKS} and \cite{FKS}.
First we consider the case when $b_{0}\left( \tau \right) \neq 0$. Also by
assumption we have $PH\leq 0$. Hence $P\left( H-\frac{1}{2}b\left( \tau
\right) ^{2}\right) ^{-1}$ is a well defined operator with norm smaller or
equal to $2/b\left( \tau \right) ^{2}.$ Therefore we have
\begin{eqnarray}
&&\left\| P\exp -ib\left( \tau \right) z\cdot \exp ic\left( \tau \right)
p_{z}\varphi \right\|  \nonumber \\
&& =\left\| P\left( H-\frac{1}{2}b\left( \tau \right) ^{2}\right)
^{-1}\left( H-\frac{1}{2}b\left( \tau \right) ^{2}\right) \exp -ib\left(
\tau \right) z\cdot \exp ic\left( \tau \right) p_{z}\varphi \right\|
\nonumber \\
&&\leq \frac{2}{b\left( \tau \right) ^{2}}\left\| \left( H-\frac{1}{2} b\left( \tau
\right) ^{2}\right) \exp -ib\left( \tau \right) z\cdot \exp ic\left( \tau \right)
p_{z}\varphi \right\| \,\,\,.  \label{zwischen}
\end{eqnarray}
Inserting the relation
\begin{eqnarray}
&&\exp -ic\left( \tau \right) p_{z}\cdot \exp ib\left( \tau \right) z\cdot
H\cdot \exp -ib\left( \tau \right) z\cdot \exp ic\left( \tau \right) p_{z}
\nonumber \\
&=&H_{0}-b\left( \tau \right) p_{z}+\frac{1}{2}b\left( \tau \right)
^{2}+V\left( \vec{x}-c\left( \tau \right) e_{z}\right)
\end{eqnarray}
into (\ref{zwischen}) yields
\begin{eqnarray}
&&\left\| P\exp -ib\left( \tau \right) z\cdot \exp ic\left( \tau \right)
p_{z}\varphi \right\|  \nonumber \\
&\leq &\frac{2}{b\left( \tau \right) ^{2}}\left\{ \left\| H_{0}\varphi
\right\| +b\left( \tau \right) \left\| p_{z}\varphi \right\| +\left\|
V\left( \vec{x}-c\left( \tau \right) e_{z}\right) \varphi \right\| \right\}
\,\,.  \label{limm}
\end{eqnarray}
Furthermore
\[
\left\| p_{z}\varphi \right\| ^{2}=\left\langle \varphi ,p_{z}^{2}\varphi
\right\rangle \leq 2\left\langle \varphi ,H_{0}\varphi \right\rangle \leq
\left\langle \varphi ,H_{0}{}^{2}\varphi \right\rangle +\left\langle \varphi
,\varphi \right\rangle ,
\]
such that
\begin{equation}
\left\| p_{z}\varphi \right\| \leq \left\| H_{0}\varphi \right\| +\left\|
\varphi \right\| \,\,\,.
\end{equation}
Finally we have that $\left\| V\left( \vec{x}-c\left( \tau \right)
e_{z}\right) \varphi \right\| $ is uniformly bounded in $E_{0}$ by Lemma 2
and therefore we may control the limit $\left| E_{0}\right| \rightarrow
\infty $ in (\ref{limm}), i.e. the r.h.s. goes as 
${\cal O}\left( \frac{1}{ \left\| E_0   \right\|  }  \right) $.
 This concludes the proof of the Main Theorem for
the case $b_{0}\left( \tau \right) \neq 0$ .

We now turn to the case when $P$ is a finite dimensional projection and $b_{0}\left(
\tau\right) ^{2}+c_{0}\left( \tau \right) ^{2}=:a_{0}^{2}\neq 0$. 
Actually by what has already been proved, it would suffice to consider the 
case $b_0(\tau)=0, c_0(\tau)\neq 0  $ only. However, we will prove the claim
for an arbitrary finite dimensional $P$ not necessarily being the projection
onto the space  spanned by the bound states  of $H$.

We start with two
preliminary considerations. First, by the Campbell-Hausdorff formula
\begin{equation}
\exp -ib\left( \tau \right) z\cdot \exp ic\left( \tau \right) p_{z}=\exp
\frac{i}{2}b\left( \tau \right) c\left( \tau \right) \cdot \exp i\left(
c\left( \tau \right) p_{z}-b\left( \tau \right) z\right) \,\,.
\end{equation}
Now there is always an $s$ such that
\begin{equation}
c\left( \tau \right) p_{z}-b\left( \tau \right) z=E_{0}a_{0}\left( z\cos
s+p_{z}\sin s\right) =E_{0}a_{0}Z(s)\,\,.
\end{equation}
Introducing the unitary operator $W(s)=\exp \frac{is}{2}\left(
p_{z}^{2}+z^{2}\right) $ we may perform a Bogoliubov transformation on $z$
\begin{equation}
W(s)\,z\,W(s)^{-1}=Z(s)\,\,\,.
\end{equation}
Secondly, let $\varphi _{n}(1\leq n\leq N)$ be an orthonormal basis for
the range of P. Then
\begin{eqnarray*}
&&\!\left\| P\exp -ib\left( \tau \right) z\cdot \exp ic\left( \tau \right)
p_{z}\varphi \right\| ^{2} \\
\qquad \qquad \qquad &=&\sum\limits_{n=1}^{N}\left| \left\langle \varphi
_{n},\exp iE_{0}a_{0}Z(s)\cdot \varphi \right\rangle \right| ^{2} \\
&=&\sum\limits_{n=1}^{N}\left| \left\langle W(s)^{-1}\varphi _{n},\exp iE_{0}a_{0}z\cdot
W(s)^{-1}\varphi \right\rangle \right| ^{2} \\ &=&\sum\limits_{n=1}^{N}\left| \int
d\vec{x}\overline{\left( W(s)^{-1}\varphi _{n}\right) (\vec{x})}\left( W(s)^{-1}\varphi
\right) (\vec{ x})\exp iE_{0}a_{0}z\right| ^{2} \;\; .
\end{eqnarray*}
Since $\overline{\left( W(s)^{-1}\varphi _{n}\right) (\vec{x})}\left( W(s)^{-1}\varphi
\right) (\vec{x})\in L^{1}\left( \mathbb{R}^{3}\right) $ the right hand side of the last
equation vanishes in the limit $\left| E_{0}\right| \rightarrow \infty $ by the
Riemann-Lebesgue theorem, which concludes the proof of the Main Theorem.$\blacksquare $

\vspace{5mm}

Now we turn to the case when $b_{0}\left( \tau \right)=c_{0}\left( \tau \right)=0$. Notice, that if we consider linearly polarized light, then  for the most
common pulse shapes like for instance a static envelope, trapezoidal 
envelope, sine-squared envelope etc. \cite{FFS} the extreme high frequency
limit, i.e. $ \omega \rightarrow \infty $ leads to  $b_{0}\left( \tau \right)
=c_{0}\left( \tau \right)=0$.  This limit is needed in order to apply the
analysis of Gavrila and coworkers \cite{HFA}, which  provides so far the
most profound ``explanation'' for the occurrence of stabilization.

We prove the Second Main Theorem:

\vspace{5mm}

\begin{theorem}
\textit{Let} $b_{0}\left( \tau \right)=c_{0}\left( \tau \right)=0$. \textit{Denote}
\begin{equation}
p(\tau)=\lim_{\left| E_{0}\right| \rightarrow \infty }\mathcal{P}\left( \psi \right)
=\left\| \left( 1-P\right) \exp\ -i\tau H_0\cdot \psi \right\| ^{2}.
\end{equation}
\textit{If H has only one bound state (i.e. if P is one dimensional), then} 
$p(\tau)>0$ \textit{for all} $\tau>0$.  

\textit{Furthermore} 
$p(\tau)<1$
\textit{(at least) for all} $\tau\in[0,\tau_\ast]$, \textit{where}
\begin{displaymath}
\tau_\ast=\pi\left[ \langle H_0\psi, H_0\psi \rangle-
\langle \psi, H_0\psi \rangle^2 \right]^{-1/2}.
\end{displaymath}
\end{theorem}

\vspace{5mm}

{\bf Proof:} We consider the survival probability 
$q(\tau)=\left|(\psi, e^{-i\tau H_0}\psi) \right|^2$ and observe that 
$p(\tau)= 1 - q(\tau) $ if there is only one bound state.
We prove that $q(\tau)<1$ for all $\tau>0$. Note that by Schwarz inequality
the bound $q(\tau)\leq 1 $ is trivial.
Now
\begin{displaymath}
\left(\psi, e^{-i\tau H_0}\psi \right)=\int_0^\infty e^{-i\lambda\tau} d\mu_\psi(\lambda)
\equiv \widehat{\mu_\psi}(\tau),
\end{displaymath}
where $\mu_\psi$ is the (nonnegative, absolute continuous) spectral measure associated
with $H_0$,
\begin{displaymath}
\mu_\psi((-\infty,\lambda])=\left\{\begin{array}{ll}
\int_{|\vec{p}|\leq\sqrt{2 \lambda}}|\widehat{\psi}(p)|^2 d\vec{p}, & \lambda\geq 0\\
0, & \lambda<0.
\end{array}\right.
\end{displaymath}
Obviously, $\int_{\mathbb{R}}d\mu_\psi(\lambda)=1$. It is well known (see e.g.
\cite{Lukacs}) that $|\widehat{\mu_\psi}(\tau)|<1$ for all $\tau>0$ 
when the measure
$\mu_\psi$ is absolutely continuous.

The second part of the theorem follows from the estimate of Pfeifer \cite{Pfeifer}.
$\blacksquare$

\vspace{5mm}

We note that due to the Paley-Wiener theorem $\widehat{\mu_\psi}(\tau)$ cannot have 
compact support. Therefore the inequality $p(\tau)<1$ must be valid for some
suitable arbitrary large $\tau>0$. On the other hand, it is well known that
$|\widehat{\mu_\psi}(\tau)|\leq C \tau^{-N}$ 
for arbitrary $N$ and all large $\tau>0$, since 
the spectrum of $H_0$ is purely transient absolute continuous 
(see for instance \cite{CFKS}). This means that in case $H$ has only one
bound state, say, the
ionization probability $p(\tau)$ will tend to one faster 
than any power inverse power of $\tau$ for $\tau \rightarrow \infty$ .

\vspace{5mm}

\textit{Example 1} The easiest pulse shape for which theorem 4 applies is
$f(t) = \cos( \omega t ) $, since then $ c_0(\tau= \frac{2 \pi n}{\omega})=
b_0(\tau= \frac{2 \pi n}{\omega}) =0$.
As a concrete  example for the potential we choose 
 the Coulomb potential
$V(\vec{x})=-1/|\vec{x}|$. The normalized wave function of the ground
state in the  
momentum representation 
is given by (see for instance \cite{BSLL})
\begin{displaymath}
\Psi(\vec{p})=\frac{\sqrt{8}}{\pi}  \frac{1}{(1+ p^2 )^2 } \;\;\; ,
\end{displaymath}
such that the survival probability in this case reads
\begin{displaymath}
q(\tau) =\left( \frac{32}{\pi} \right)^2 
\left| \int\limits_{0}^{\infty} dp \, p^2 
\frac{e^{-i \tau \frac{p^2}{2} } }{(1+ p^2 )^4 }  \right|^2 =  
\frac{64}{\pi} 
\left| U(\frac{3}{2}, - \frac{3}{2}; \frac{i \tau}{2}   )   \right|^2 
\;\;\; .
\end{displaymath}
Here $U(a,b;z)$ denotes a confluent hypergeometric function 
(see for instance \cite{Abr}). So for 
typical  sub-picosecond pulses we obtain for instance $q(\tau= 400 a.u.)= 
2.45 \; 10^{-6}$
and  $q(\tau=1000 a.u.)= 1.63 \; 10^{-7}$. Essential is here to note that the 
survival probability is always non-vanishing and monotonically decreasing
in $\tau$ (see figure 1).

\textit{Example 2} We now take the potential 
to be  the  attracting point interaction, often also called
the delta potential in three dimensions, 
(see e.g.
\cite{Alb}) with coupling constant $\alpha>0$. This potential  has the 
virtue that it possess only one eigenstate
\begin{displaymath}
\Psi(\vec{x})= \sqrt{ \frac{\alpha}{2 \pi }}  
\frac{e^{-\alpha|\vec{x}|}}{|\vec{x}|}
\end{displaymath}
with energy $-(\alpha)^2$. In the momentum representation the wave 
function is given by
\begin{displaymath}
\Psi(\vec{p})= \frac{\sqrt{\alpha}}{\pi} \frac{1}
{\left(\alpha^2+ p^2 \right) }
\end{displaymath}
such that the survival probability turns out to be
\begin{displaymath}
q_\alpha(\tau) = \frac{16 \alpha^2}{\pi^2 }  
\left| \int\limits_{0}^{\infty} dp \, p^2 
\frac{e^{-i \tau \frac{p^2}{2} } }{(  \alpha ^2+ p^2 )^2 }  
\right|^2 =  
\frac{1}{\pi} 
\left| U(\frac{3}{2},  \frac{1}{2}; \frac{i \tau  \alpha^2}{2}   )   \right|^2 
\;\;\; .
\end{displaymath}
As figure 2 illustrates, the survival probability 
decreases monotonically with increasing $\alpha$ for fixed 
pulse duration $\tau$. The figure also shows, that for increasing 
$\tau$ the survival probability decreases. 

We may assume that the Hydrogen atom behaves with respect
to the energy variation qualitatively  the same way as the 
point interaction. Then
this example indicates that  one should  expect
that for sufficiently high Rydberg states the survival probability 
$q(\tau)$ will  be
sufficiently close to 1 even for  times $\tau\approx 1$ ps.

\par
\textbf{Conclusion}
We have investigated  the ionization probability in the  extreme intensity
limit for three different situations. The first analysis presumes that 
the classical momentum transfer $b_0(\tau)$ is non-vanishing and allows
essentially all common potentials. Since the condition  $b_0(\tau)\neq 0 $
excludes a wide range of possible pulses, we also studied separately
a situation for which we only demand that not both the classical momentum
transfer $b_0(\tau) $ and the total classical momentum transfer
$c_0(\tau) $ vanish simultaneously. In addition we have to demand for this
case that potential only possess a finite number of bound states. This is
similar to the situation in many numerical calculations, in which one 
is also forced to introduce a cut-off at some level when projecting onto
bound states. In both cases we find that the ionization probability ${\cal P}$ 
for any bound state of the Hamiltonian $H = H_0 + V$ for the field amplitude
$E_0$ going to infinity is going to one. 
This excludes in our opinion the possibility
of stabilization for these situations, apart from window stabilization.

Finally, we considered the situation in which $b_0(\tau) =c_0(\tau)=0 $ and
find indeed the possibility of stabilization. For the most common pulses, 
which 
involve linearly polarized light, this case corresponds to the high
frequency limit of Gavrila and coworkers \cite{HFA}. We conclude that 
our analysis is consistent with the ``high frequency picture'' and that
stabilization is only to be expected in this latter case.

\input epsf
\begin{figure}[hbt]
\caption{Survival probability $q(\tau(a.u.))$ after the time $ \tau$ 
for the  ground state of the Hydrogen atom under the free time evolution.}
\end{figure}

\begin{figure}[hbt]
\caption{Survival probability $q_\alpha(\tau(a.u.))$ for fixed pulse
duration $ \tau=200 a.u. $ dotted line, $ \tau=400 a.u. $ 
dashed line, $ \tau=1000 a.u. $ solid line,  
for the bound state of the three dimensional delta potential
 under the free time evolution as a function of the coupling $\alpha$.}
\end{figure}

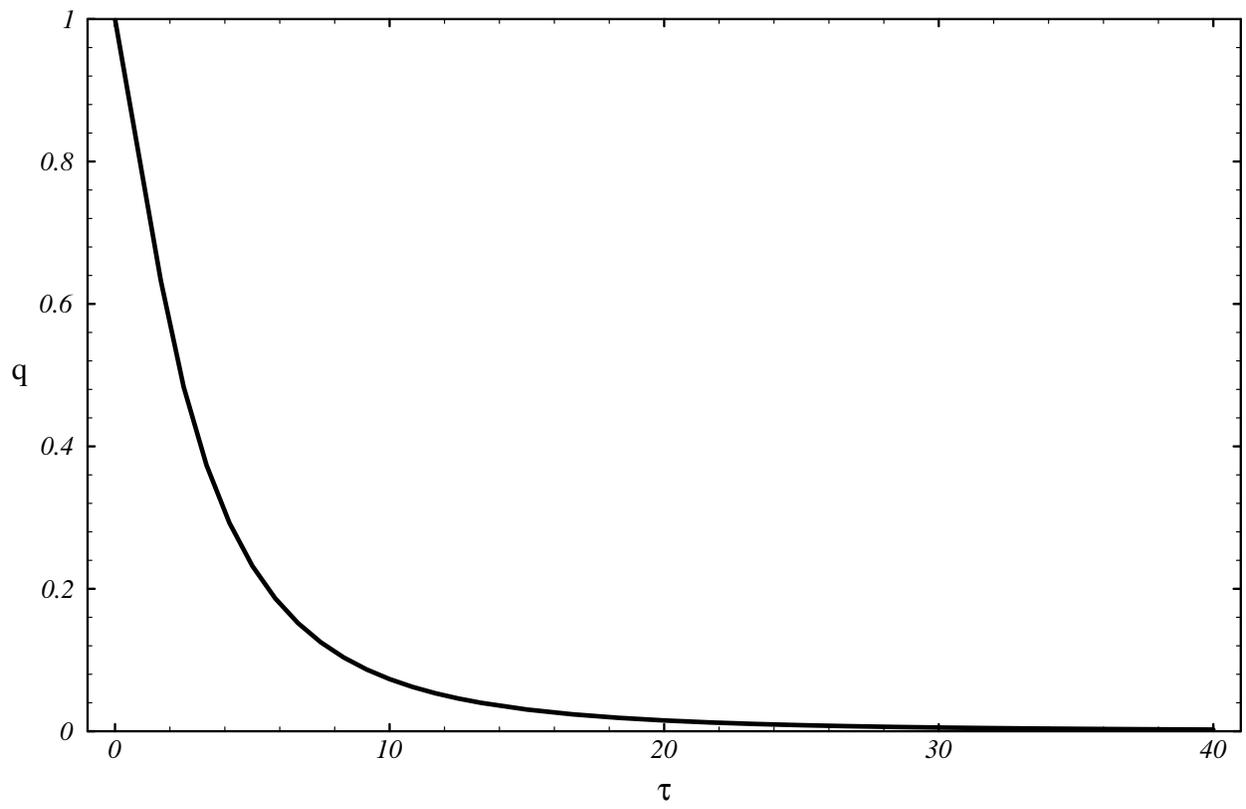
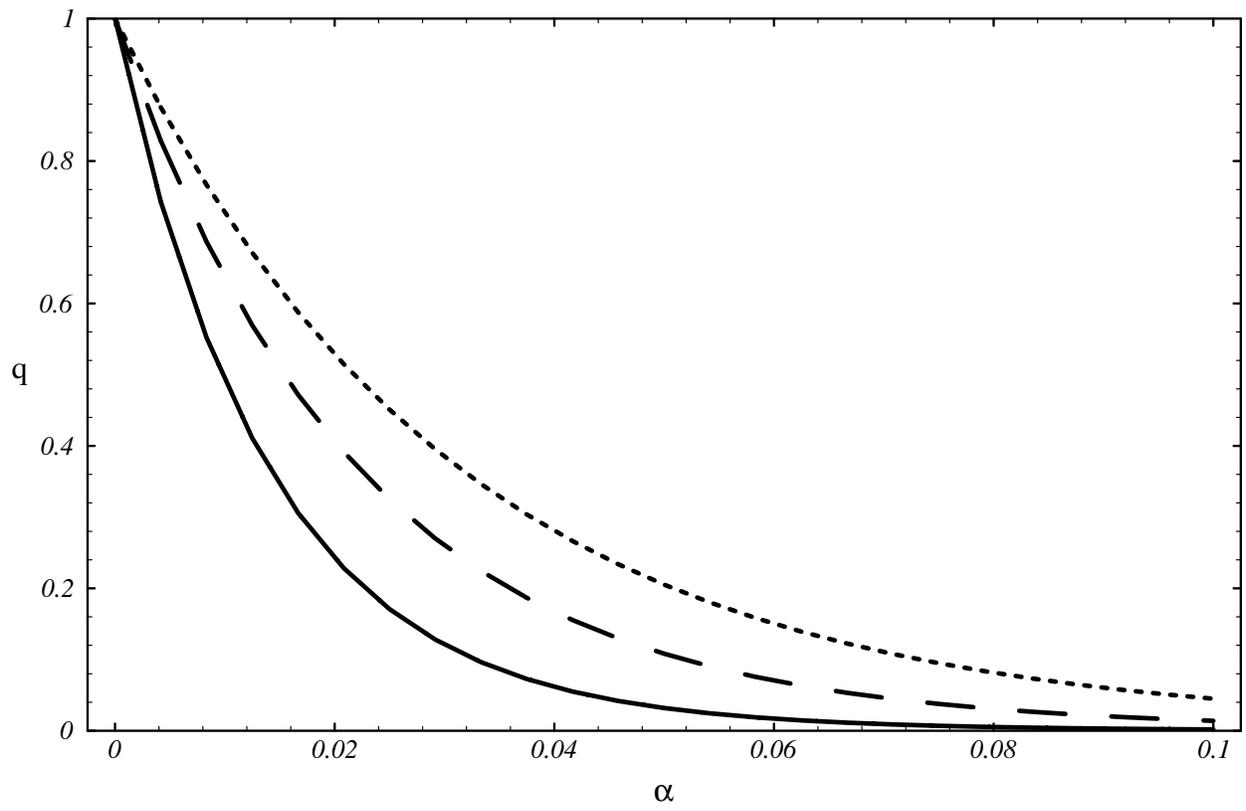
\end{document}